\documentclass{aastex61}
\usepackage{graphicx}
\usepackage{amsmath,bm}

\usepackage{natbib}
\shorttitle{X-class flares on 2017 September 06}
\shortauthors{Mitra et al.}

\begin{document}

\title{Successive flux rope eruptions from $\delta$-Sunspots region of NOAA 12673 and associated X-class eruptive flares on 2017 September 6}

\correspondingauthor{Prabir K. Mitra}
\email{prabir@prl.res.in}

\author{Prabir K. Mitra}
\affiliation{Udaipur Solar Observatory, Physical Research Laboratory, Udaipur 313 001, India}

\author{Bhuwan Joshi}
\affiliation{Udaipur Solar Observatory, Physical Research Laboratory, Udaipur 313 001, India}

\author{Avijeet Prasad}
\affiliation{Udaipur Solar Observatory, Physical Research Laboratory, Udaipur 313 001, India}

\author{Astrid M. Veronig}
\affiliation{Institute of Physics \& Kanzelh\"ohe Observatory, University of Graz, Universit\"{a}tsplatz 5, A-8010 Graz, Austria}

\author{R. Bhattacharyya}
\affiliation{Udaipur Solar Observatory, Physical Research Laboratory, Udaipur 313 001, India}

\begin{abstract}
In this paper, we present a multi-wavelength analysis of two X-class solar eruptive flares of classes X2.2 and X9.3 that occurred in the sigmoidal active region
NOAA 12673 on 2017 September 6, by combining observations of Atmospheric
Imaging Assembly and Helioseismic Magnetic Imager instruments
on board the Solar Dynamics Observatory. On the day of the reported activity, the photospheric structure of the active region displayed a very complex network of $\delta$-sunspots that gave rise to the formation of a coronal sigmoid observed in the hot EUV channels. Both X-class flares initiated from the core of the sigmoid sequentially within an interval of $\sim$3 hours and progressed as a single ``sigmoid--to--arcade'' event. Differential emission measure analysis reveals strong heating of plasma at the core of the active region right from the pre-flare phase which further intensified and spatially expanded during each event. The identification of a pre-existing magnetic null by non-force-free-field modeling of the coronal magnetic fields at the location of early flare brightenings and remote faint ribbon-like structures during the pre-flare phase, which were magnetically connected with the core region, provide support for the breakout model of solar eruption. The magnetic extrapolations also reveal flux rope structures prior to both flares which are subsequently supported by the observations of the eruption of hot EUV channels. The second X-class flare diverged from the standard flare scenario in the evolution of two sets of flare ribbons, that are spatially well separated,
providing firm evidence of magnetic reconnections at two coronal heights. 
\end{abstract}


\section{Introduction}
Solar eruptive flares and associated Coronal Mass Ejections (CMEs) produce large-scale changes in coronal magnetic field configurations. The magnetic and plasma structures of Earth directed CMEs have regularly been recorded by in-situ measurements at 1 AU and these are known to drive hazardous effects at near-Earth environment. To probe the exact magnetic configuration of flare producing active regions during the pre-eruption phase and its role in driving the subsequent eruptive phenomena have been among the major objectives of contemporary research in solar physics.

The standard flare model, also known as CSHKP model \citep{Carmichael1964, Sturrock1966, Hirayama1974, Kopp1976}, recognizes that the initiation of a solar eruptive flare is intrinsically related to the dynamical activation of a filament (or prominence) by some kind of instability in the system. Once the filament attains eruptive motions, large-scale magnetic reconnection sets in underneath it in a vertical current sheet that accelerates the particles to relativistic speeds and heats the plasma to million Kelvin (MK) temperatures. It is generally accepted that the energy source for the observed eruptive phenomena is derived from the free energy contained in sheared or twisted magnetic fields \citep{Forbes2000}. Although the standard flare model takes into account most of the commonly observed phenomena, such as, flare ribbons, looptop and footpoint sources, cusp structure following the passage of the filament, post-flare loop arcade etc. \citep[e.g., see review by][]{Benz2017,Fletcher2011}; it remains silent regarding the mechanism that triggers the filament eruption at the first place. We should also note that despite the general applicability of the standard flare model, many CME/eruptive flares exhibit features that deviate from this scenario \citep[e.g., see][]{Joshi2012}.

From theoretical point of view, the structure of an erupting filament or filament channel is attributed to a magnetic flux rope \citep[e.g., see][]{Gibson2006} whose exact topology and formation process remain unclear and debatable. There are two basic classes of models that describe the magnetic configuration of the system initially: sheared arcade and twisted flux rope. A sheared magnetic arcade extends along the polarity inversion line (PIL) while the flux rope is thought to lie above the inversion line. Considering these initial configurations, the onset of eruption can be attributed to either loss of equilibrium/ideal instability or magnetic reconnection \citep[see reviews by][]{Klimchuk2001,Forbes2006,Vrsnak2008,Chen2011,Green2018}. In general, the pre-existence of the flux rope is a prerequisite for ideal models. On the other hand, reconnection models can work with either a sheared arcade or a twisted flux rope. In observations, the coronal sigmoid \citep[e.g., see][]{Manoharan1996,Hudson1998} and/or erupting hot EUV channels \citep[e.g., see][]{Patsourakos2013,Cheng2016} are thought to provide strong support toward flux rope scenario. Thorough multi-wavelength case studies of major solar eruptions are extremely useful to probe whether the flux rope signatures pre-exist or develop on the fly during the pre- or early phases of solar flares.

To illustrate the eruptive flares in complex active regions (ARs), we divide the underlying magnetic configuration in two parts: core and envelope fields \citep{Moore2001}. Core fields denote the low-lying magnetic flux system, close to the PIL, and it is assumed that a major part of the stored magnetic energy lies in this region. The envelope refers to outer, large-scale overlying flux system. For a successful eruption, the stressed core field must destabilize and eventually erupt through the barrier of the overlying flux system to generate a CME. Two representative models of solar eruptions-- tether-cutting and breakout-- relies on the magnetic reconnection before the flare (and/or CME onset) that trigger the eruption of the core fields. However, the early reconnection proceeds quite differently in the two models. The pre-flare configuration in ``tether--cutting model" comprises a single, highly sheared, bipolar AR, with the earliest reconnection (i.e., triggering process) taking place deep in the sheared core fields \citep{Moore1992, Moore2001}. On the other hand, the ``breakout model" involves a multipolar topology, containing one or more pre-existing coronal null points. In this case, the CME onset is triggered by reconnection occurring well above the core region \citep{Antiochos1999}. The basic reconnection conditions for both the models give rise to differences in the sequence and location of the pre-flare brightenings. In tether-cutting model, all pre-flare activity resides beneath the flare reconnection site. In contrast, the breakout model causes both core and remote brightenings prior to flare/CME initiation \citep[e.g., see][]{Sterling2001}. A thorough investigation of the sites of pre-flare brightenings with respect to the location of flux rope/filament along with coronal null-points is essential to understand the triggering process of solar eruptions. However, after the successful trigger, once the flux rope attains eruptive expansion, the standard flare reconnection occurs in a vertical current sheet that is formed underneath the flux rope and this scenario is common to all the models of eruptive flares \citep[e.g., see][]{Veronig2006, Liu2008,Joshi2013, Joshi2016}.

In this paper, we present comprehensive multi-wavelength analysis of two major solar eruptive flares of classes X2.2 and X9.3 which occurred on 2017 September 6 in the active region NOAA 12673. Interestingly, the two flares initiated from the same location of the sigmoidal active region and occurred within an interval of $\sim$3 hours. We investigate the processes associated with flare initiation and subsequent eruption by utilizing multi-channel, high resolution observations from the Atmospheric Imaging Assembly \citep[AIA;][]{Lemen2012} on board the Solar Dynamics Observatory \citep[SDO;][]{Pesnell2012}. In particular, the study of the distribution of hot plasma within the sigmoid at various phases by Differential Emission Measure (DEM) technique provide important insights about the plasma heating caused due to energy deposition by accelerated particles. We have further undertaken a comprehensive study of the photospheric magnetic structures; thanks to high resolution LOS magnetograms from the Helioseismic and Magnetic Imager \citep[HMI;][]{Schou2012}. 
Further, we use the photospheric vector magnetograms from HMI as an input to a non-force-free-field magnetic field extrapolation model \citep{Hu2008b}. Such extrapolations have been recently used to successfully model the magnetohydrodynamic evolution of flaring active regions \citep[see][]{Prasad2018}.
Multi-wavelength analysis in combination with coronal magnetic field modeling have clarified important aspects of solar eruptions, such as, triggering of flux rope eruption and its early evolution in the low corona which are examined in terms of scope and limitations of the standard flare model. Section 2 provides a brief account of the observational data and analysis techniques. In Section 3, we present multi-wavelength analysis and derive the observational results on the basis of comprehensive measurements taken at photospheric, choromspheric and coronal levels. In Section 4, we compare the chromospheric and coronal observations of different flare associated features with magnetic topology and configurations obtained with non-force-free-field modeling of the active region corona. We discuss and interpret our results in Section 5.

\section{Observations and Data Analysis Techniques}
In this paper, we use observational data from the Atmospheric Imaging Assembly \citep[AIA;][]{Lemen2012} and the Helioseismic and Magnetic Imager \citep[HMI;][]{Schou2012} on board the Solar Dynamics Observatory \citep[SDO;][]{Pesnell2012}. AIA observes the Sun in seven EUV filters (94 \AA , 131 \AA , 171 \AA , 193 \AA , 211 \AA , 304 \AA\ and 335 \AA ), two UV filters (1600 \AA\ and 1700 \AA ) and one white light filter (4500 \AA ). AIA produces 4096$\times$4096 pixel full disk solar images at a spatial resolution of $0''.6$ pixel$^{-1}$ and temporal cadence of 12 s for EUV filters, 24 s for UV filters, and 3600 s for the white light filter.

Differential Emission Measure (DEM) analysis provides insights of multi-thermal processes associated with energy release in solar eruptive phenomena. Observations from the six coronal EUV filters of AIA (94 \AA , 131 \AA , 171 \AA , 193 \AA , 211 \AA , and 335 \AA ) are used for computation of DEM. A brief account of theoretical background for DEM along with its analysis technique is given in Section 3.4.

For studying the photospheric structures associated with the active region NOAA 12673, we have used HMI intensity and line-of-sight (LOS) magnetogram images. HMI produces full disk LOS intensity (continuum) and magnetogram images of 4096$\times$4096 pixel at spatial resolution of $0''.5$ pixel$^{-1}$ and 45 s temporal cadence. In order to have exact spatial alignment between AIA and HMI images, we have processed the HMI images by SSW routine \textit{hmi\_prep.pro} which converts spatial resolution of HMI images from $0''.5$ pixel$^{-1}$ to $0''.6$ pixel$^{-1}$ and resolves the $\sim$180$^\circ$ roll angle of level 1 images.

Coronal magnetic field lines are extrapolated by using the `hmi.sharp$\_$cea$\_$720s' series from SDO/HMI which takes full disk vector magnetograms at a spatial resolution ${0''.5}$ pixel${^{-1}}$ and temporal cadence of 720 s. The magnetograms in this dataset are first remapped on to a Lambert cylindrical equal-area (CEA) projection and then transformed into the heliographic coordinates \citep{Gary1990} which is essential for extrapolations in Cartesian geometry. In this case, the magnetogram represents a cut-out image of the active region of dimensions 688$\times$448 pixels which corresponds to a physical size of $\sim$250 and $\sim$160 Mm, respectively, on the Sun. Extrapolations were done up to an extent of 448 pixels vertically which translates to a height of $\sim$160 Mm from the photosphere. For visualizing the modeled field lines, we have used Visualization and Analysis Platform for Ocean, Atmosphere, and Solar Researchers \citep[VAPOR\footnote{\url{https://www.vapor.ucar.edu/}};][]{Clyne2007} software which produces an interactive 3D visualization environment.

Coronal mass ejections associated with the X-class flares were observed by the C2 and C3 of the Large Angle and Spectrometric Coronagraph \citep[LASCO;][]{Brueckner1995} on board the Solar and Heliospheric Observatory \citep[SOHO;][]{Domingo1995}. C2 and C3 are white light coronagraphs imaging from 1.5 to 6 R$_\odot$ and from 3.7 to 30 R$_\odot$, respectively.

\section{Multi-wavelength analysis and Results}

\subsection{Event Overview}
The investigation carried out in this paper corresponds to the observation of the active region NOAA 12673 on 2017 September 6 from 08:00 UT to 14:00 UT. In this period, the active region produced two X-class flares; both associated with CME.
\begin{figure}
\plotone{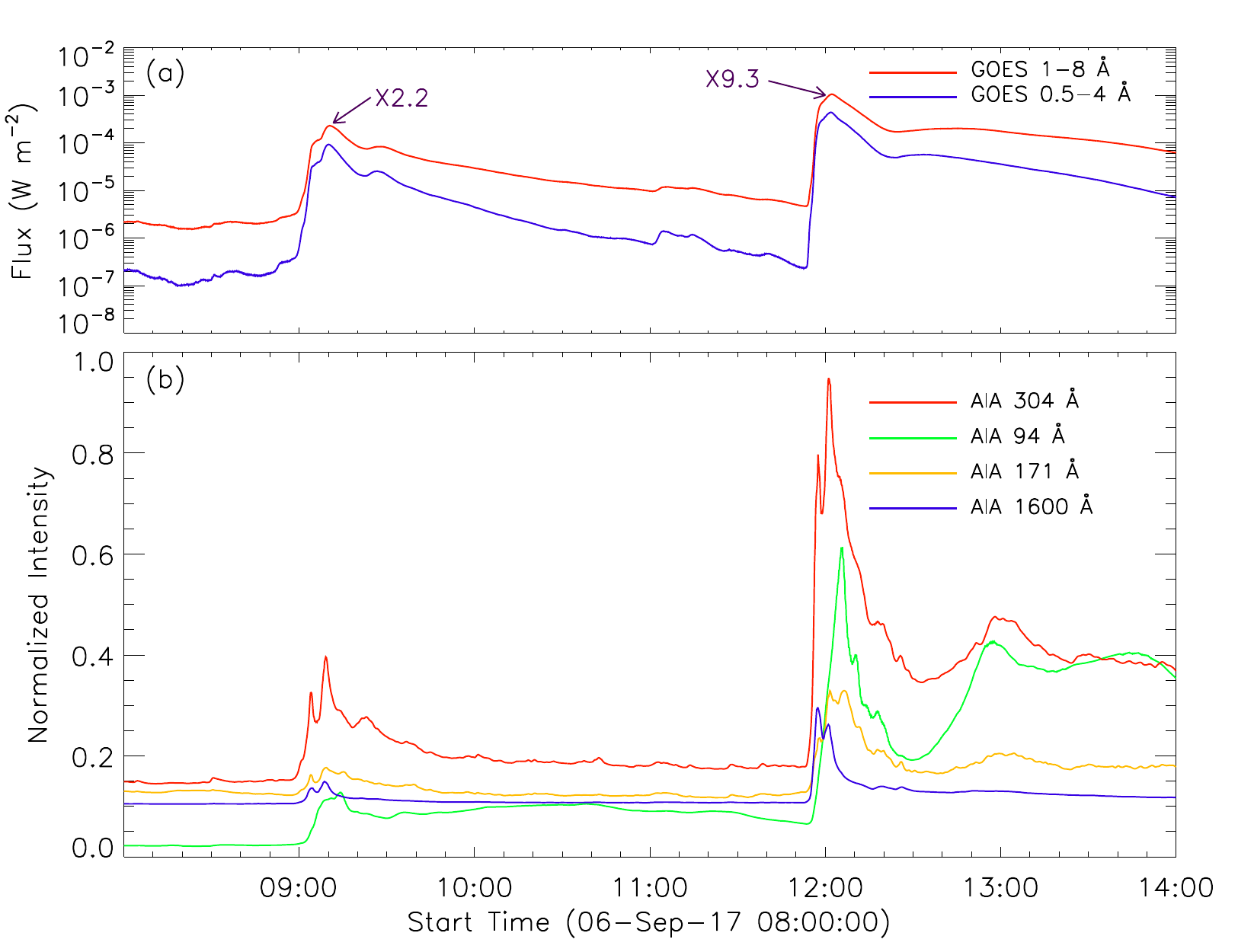}
\caption{Panel (a): GOES X-ray flux profiles in 1--8 \AA\ and 0.5--4 \AA\ from 08:00 UT to 14:00 UT on 2017 September 6, showing the occurrence of two X-class flares of intensities X2.2 and X9.3, respectively. Panel (b): AIA lightcurves normalized by the peak intensity of the respective AIA filters. For clear visualization, the AIA lightcurves have been scaled by factors of 1, 0.65, 0.33, and 0.33 for 304 \AA , 94 \AA , 171 \AA , and 1600 \AA\ channels, respectively.}
\end{figure}
In Figure 1, we compare the soft X-ray (SXR) lightcurves from GOES (Figure 1(a)) with (E)UV time-profiles from AIA (Figure 1(b)). Lightcurves at all the energies suggest occurrence of the impulsive phase of the first flare (X2.2) between $\sim$09:00--09:10 UT. After $\sim$09:16 UT, the X ray flux of either channels slowly decreased up to $\sim$11:55 UT and thereafter rapidly increased, indicating the sudden initiation of the second flare (X9.3). The abrupt rise in X-ray flux is accompanied by rapid build up of (E)UV emission.  After $\sim$12:00 UT, the X-ray flux underwent fast decay until $\sim$12:22 UT, which was followed by an extended phase during which the flux initially increased and then declined very gradually. Notably, this two-step evolution of the second flare's decay phase is more prominent in the EUV lightcurves (cf. Figures 1(a) and (b)). A summary of these two X-class flares is given in Table 1.

\begin{deluxetable*}{p{1cm}p{2cm}p{1.5cm}p{2cm}p{1.5cm}p{5cm}}
\tablenum{1}
\tablecaption{Summary of the two X class flares occurred on 2017 September 6}
\tablewidth{0.1pt}
\tablehead{
\colhead{Sr.} & \colhead{Flare} & \colhead{} & \colhead{Time} &
 \colhead{} & \colhead{Location of the} \\
\cline{3-5}
\colhead{No.} & \colhead{Class} & \colhead{Start} & \colhead{Peak} &
 \colhead{End} & \colhead{center of the sigmoid}
}

\startdata
\ \ \ 1 & \ \ \ \ \ \ X2.2 & \ 08:59 UT & \ \ \ \ 09:10 UT & \ \ 09:36 UT & \ \ \ \ \ \ \ \ \ (x, y) $\approx$ ($510''$, $-230''$) \\
\ \ \ 2 & \ \ \ \ \ \ X9.3  &\ 11:53 UT & \ \ \ \ 12:02 UT & \ \ 13:09 UT & \ \ \ \ \ \ \ \ \ (x, y) $\approx$ ($530''$, $-250''$) \\
\enddata
\end{deluxetable*}

According to the Solar Eruptive Event Detection System (SEEDS) CME catalog\footnote{\url{http://spaceweather.gmu.edu/seeds/monthly.php?a=2017&b=09}}, both X-class flares were associated with CMEs.  The CME corresponding to the X2.2 flare was detected at 11:00 UT at 5.02 R$_\odot$ (half-max lead). This relatively narrow CME (angular width 44$^\circ$) was traveling along PA of 255$^\circ$ with linear fit speed 279 km s$^{-1}$ in the field of view of LASCO C2 (top panels of Figure 2). The CME corresponding to the X9.3 flare was detected at 12:36 UT at 3.50 R$_\odot$ (half-max lead) which had an angular width of 145$^\circ$ and linear fit speed 505 km s$^{-1}$ along PA 241$^\circ$ (bottom panels of Figure 2).

\begin{figure}
\epsscale{0.8}
\plotone{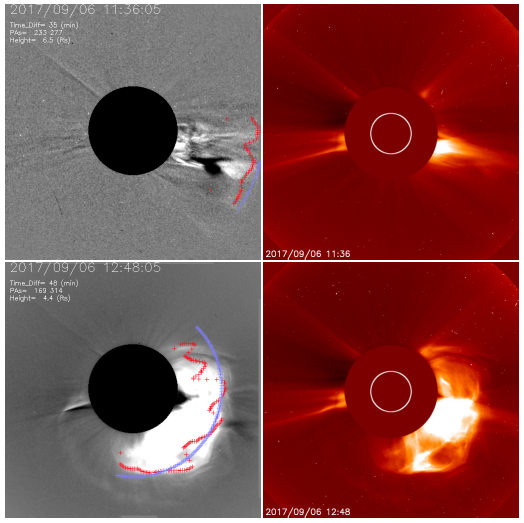}
\caption{Top panels: Running difference (left) and direct (right) images obtained from LASCO C2 coronagraph showing the CME associated with the X2.2 flare. Bottom panels: Running difference (left) and direct (right) images obtained from LASCO C2 coronagraph showing the CME associated with the X9.3 flare. The blue and red colored lines on the difference images indicate the approximate position of the leading edge created by using two different techniques.}
\end{figure}
\subsection{X2.2 flare}
\subsubsection{Flare Evolution}
Figure 3 displays a series of AIA 94 \AA\ (panels (a)--(f)) as well as AIA 304 \AA\ (panels (g)--(l)) images showing the time evolution of the X2.2 flare. The overall activity site appeared to form an inverted `S -shaped' extended region in the corona, i.e., a coronal sigmoid \citep{Rust1996, Manoharan1996}. As expected, this sigmoidal structure was most prominent and distinct during the pre-flare stages (indicated by black dotted curved lines in Figures 3(b) and (h)). Although, we noticed a few compact and short lived structures in the sigmoidal region, they did not seem to contribute much toward the overall brightness variation in the pre-flare phase. The only stand out and consistent feature, observed from long before the initiation of the flare, was a small loop which is shown in the box in Figure 3(b). Flare initiation occurred at $\sim$09:00 UT when this loop like structure started broadening up. This was followed by the activation of a filament in the middle of the sigmoid shaped in the form of a `question mark' (`?') notation (see Figure 3(c)). This filament emitted intensely during the peak phase of the flare (Figures 3(c)--(d) and in the respective zoomed cut-outs). The comparison of spatial distribution of the filament with photospheric magnetic contours reveals the filament to lie over the photospheric PIL (the PIL has been indicated by the tiny blue arrows in the cut-out of Figure 3(c)). The filament activation ceased at $\sim$09:15 UT and the flare emission started to decline thereafter. Although, the activated filament did not show any further evolution in the corona following the flare maximum, the signatures of plasma eruption are simultaneously noted from the overlying coronal environment (marked by white arrows in Figure 3(e)). The eruption scenario is further discussed in section 3.2.2.

The observation in AIA 304 \AA\ reveals some interesting aspects of energy release processes in chromospheric layer. A very localized bright spot can be found in AIA 304 \AA\ images before the flare (indicated by an arrow in Figure 3(h)). We recall emergence of a bright compact loop structure in AIA 94 \AA\ images (region inside the box in Figure 3(b)) prior to the X2.2 flare at the same region (cf. Figures 3(b) and (h)). This spatial and temporal correlation between coronal and chromospheric activities suggest a magnetic coupling between the chromospheric and coronal layers during small-scale pre-flare processes. The peak phase of the flare in AIA 304 \AA\ emission in characterized by two distinct peaks at $\sim$09:04 UT and $\sim$09:09 UT (cf. Figure 1(b)). The corresponding images (Figures 3(i)--(j)) clearly show intense emission from regions close to the activated filament while the ribbon structures remained short and not so well separated during the maximum phase (Figures 3(i) and (j)) and beyond. AIA 304 \AA\ images reveal a bright, closed loop-like structure in the decay phase, at the middle of the sigmoid (Figure 3(l)) which possibly resulted from cooling of a hot loop observed in the AIA 94 \AA\ filter during early flare stages (Figures 3(c)--(d)).
\begin{figure}
\epsscale{1.1}
\plotone{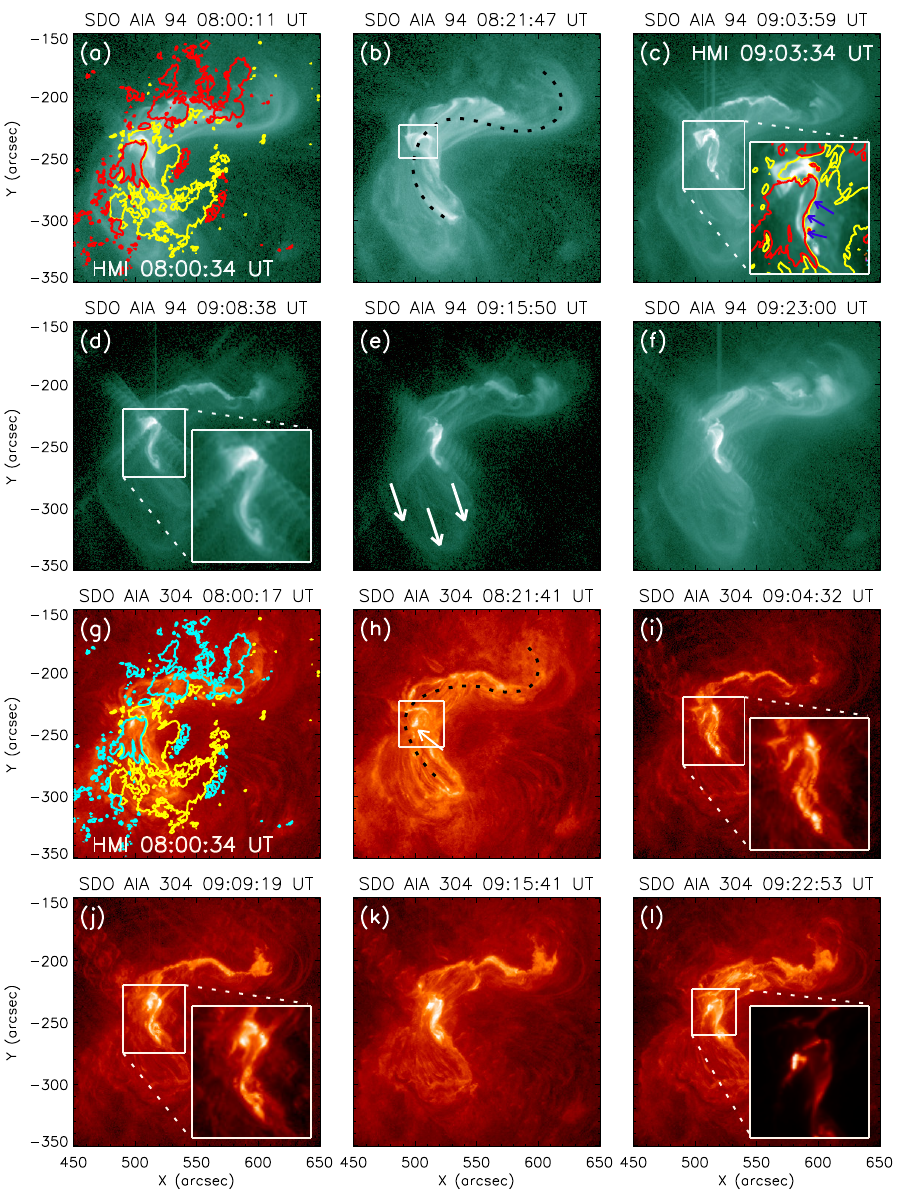}
\caption{Sequence of AIA 94 \AA\ images (panels (a)--(f)) and AIA 304 \AA\ images (panels (g)--(l)) showing the morphological evolution of the X2.2 flare. Insets in panels (c), (d), (i), (j), and (l) show zoomed-in images of the selected regions shown in smaller boxes in the corresponding panels. Contours of HMI LOS magnetograms are overplotted on selected AIA images (in panels (a), (c), and (g)) at levels of ${\pm250\ G}$. Yellow contours refer to positive polarity while red and blue contours refer to negative polarity.}
\end{figure}

\subsubsection{Plasma Eruption}
During the X2.2 flare, eruption of plasma was observed in different AIA channels. Interestingly, plasma eruption was more prominent in AIA filters sensitive to high temperatures. We have produced a series of AIA 94 \AA\ running difference images for clear display of plasma ejection during this flare (Figures 4(a)--(d)). We readily observe plasma eruption toward south-western direction from the sigmoid (black arrows in Figures 4(a)--(d) indicate the direction of erupting plasma). The amount and direction of plasma flow inferred from AIA images confirm its association with the CME detected by LASCO at 11:00 UT (top panels of Figure 2).

\subsection{X9.3 flare}
\subsubsection{Flare Evolution}
After the first flare, the active region was quiet for about two and a half hours and then it produced the largest flare of this decade of class X9.3. Figure 5 displays different phases of the flare in AIA 94 \AA\ images. In the pre-flare phase (Figures 5(a)--(c) and Figure 1(b)) we do not find any significant variation in morphology and intensity of emission within the sigmoid. With the flare onset (Figure 5(d)), we find the filament which had undergone through activation phase during the earlier flare of class X2.2, again intensified and appeared brighter than in rest of the sigmoid. The AIA 94 \AA\ emission peaked at $\sim$12:05 UT (Figure 1(b)). Images taken during the impulsive and peak phases, following the eruption of the heated filament, reveal enormous increase of brightness in the flaring area (Figures 5(e)--(f)). After $\sim$12:05 UT, the flare entered in the gradual phase where its intensity started to decrease both in GOES channels and AIA lightcurves (cf. Figure 1(b)). During this time, we identified an interesting feature occurring in the upper-right portion of the sigmoid (i.e., the elbow region; indicated by the boxes drawn with dashed lines in Figures 5(g)--(i)). The elbow exhibited a semi-circular extension at the very northern end of the sigmoid from where plasma was observed to erupt during the declining phase of the flare. By 12:30 UT, intensity of the emission from the sigmoid decreased significantly and we find a local minimum in the AIA 94 \AA\ lightcurve at $\sim$12:30 UT (which can also be noticed in other AIA lightcurves). After this time, bright coronal loops in the northern part of the sigmoid started to expand causing the intensity to increase in the AIA lightcurves again. The sigmoid with bright coronal loops in the northern part is shown in Figures 5(j)--(l). In Figure 5(l), from the HMI overplotted contours, one can easily infer that the post-flare loop arcade connected the northern negative magnetic polarity region to the central positive magnetic polarity region.

\begin{figure}
\epsscale{1.2}
\plotone{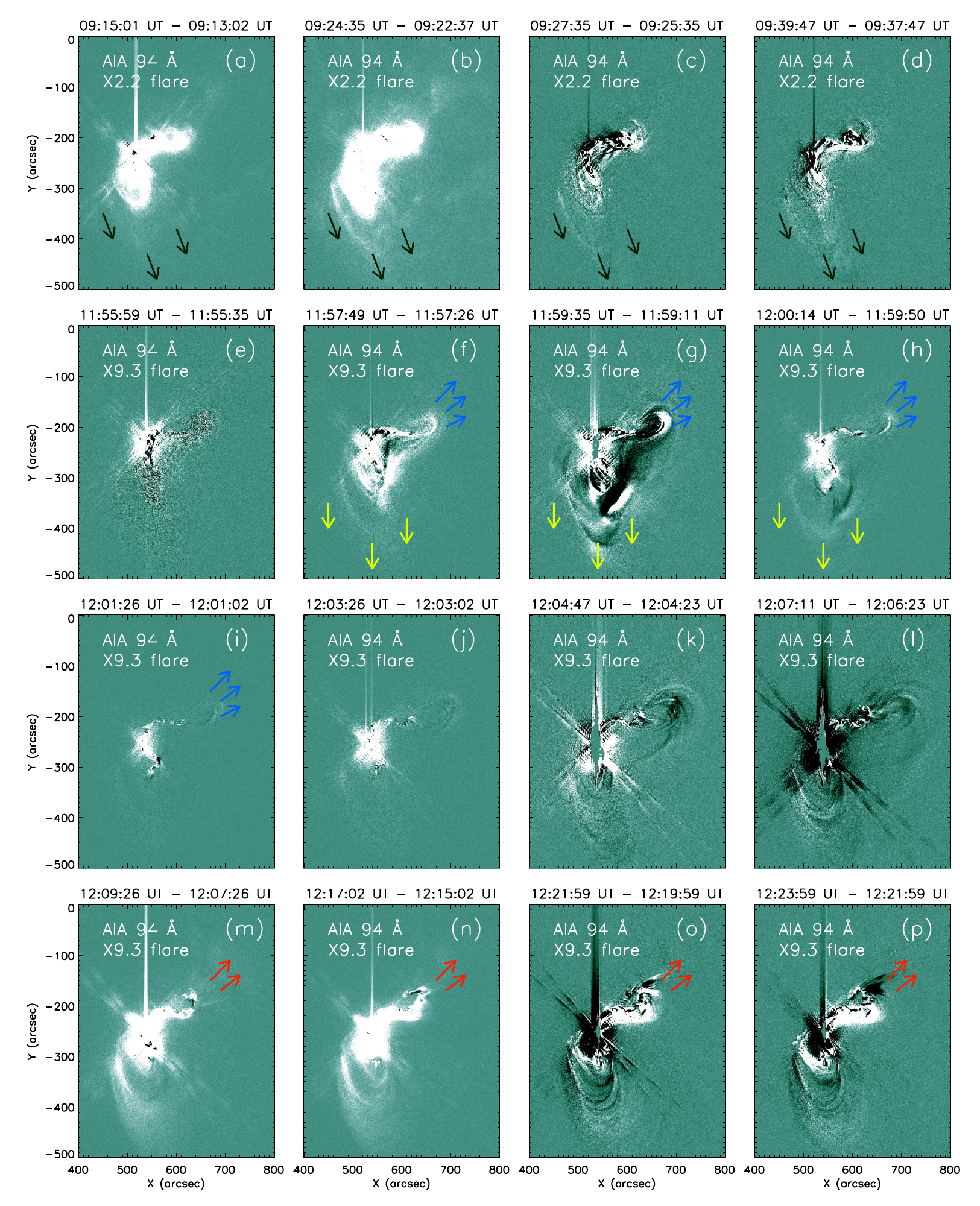}
\caption{Sequence of AIA 94 \AA\ running difference images showing different phases and directions of plasma eruption from the active region during the X2.2 (panels (a)--(d)) and the X9.3 (panels (e)--(p)) flares. Arrows in different panels indicate direction of motion of erupting plasma.}
\end{figure}

\begin{figure}
\epsscale{1.1}
\plotone{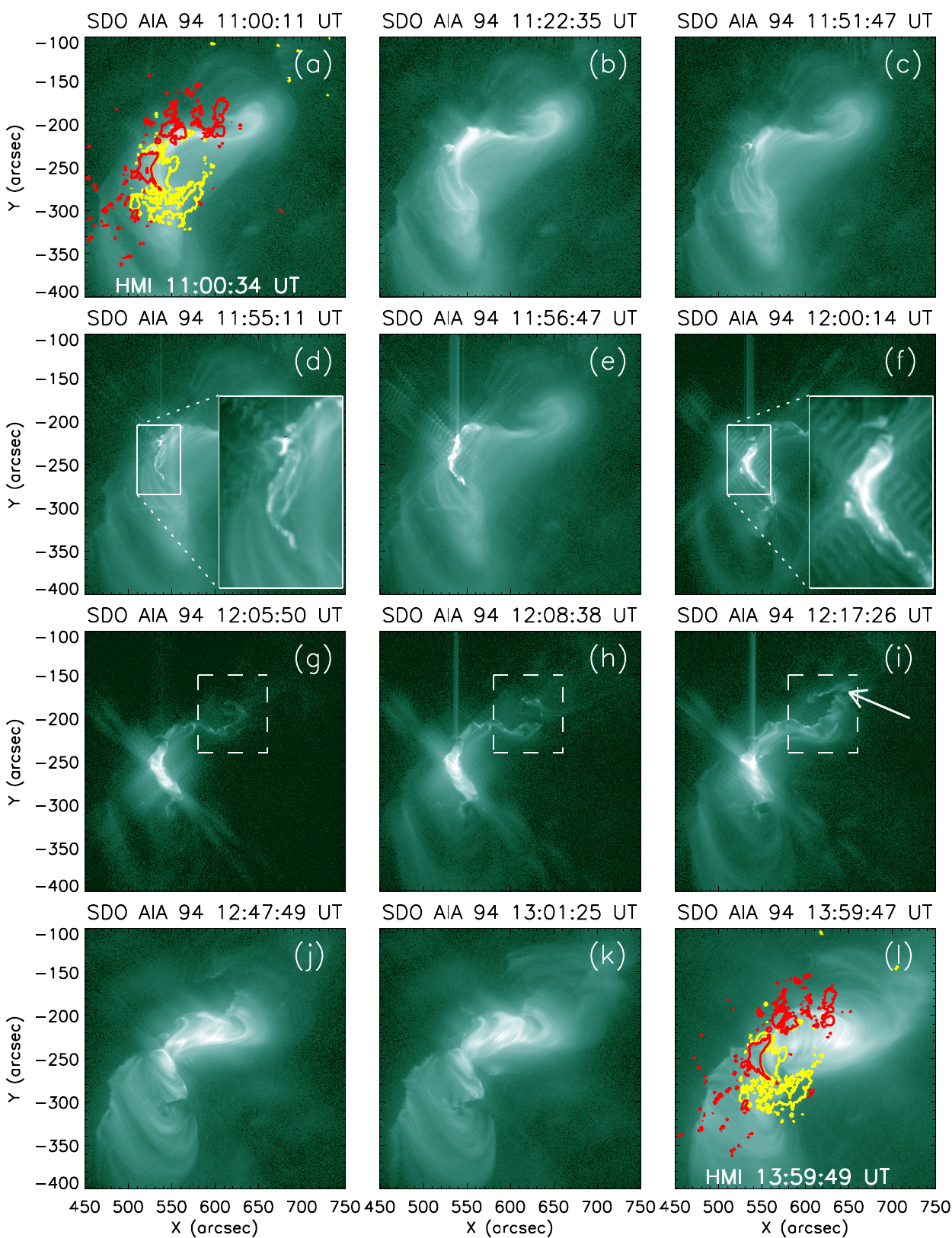}
\caption{Sequence of AIA 94 \AA\ images showing the evolution of the X9.3 flare. The inset in panel (d) shows a zoom of the central region of the sigmoid from where the flare was initiated. The inset in panel (f) shows zoomed-in image of the brightest region during the peak emission of the X9.3 flare. Regions inside the dashed boxes in panels (g)--(i) show formation of a semi-circular arc in the northern end of the active region. The arrow in panel (i) indicate eruption of plasma from the semi-circular arc region. Contour levels of HMI LOS magnetogram in panels (a) and (l) are ${\pm400\ G}$ where yellow and red contours refer to positive and negative polarity, respectively.}
\end{figure}

Figure 6 contains a series of AIA 304 \AA\ images displaying the different phases of the X9.3 flare. Before $\sim$11:53 UT (Figures 6(a)--(c)), the active region did not show much activity and only a few relatively less significant, compact, and localized brightenings were observed. The filament lying in the middle portion of the sigmoid started to brighten up from $\sim$11:53 UT which resulted in the onset of the impulsive phase of the X9.3 flare. The filament, after going through a brief phase of rapid expansion at around 11:55 UT, eventually erupted at $\sim$11:56 UT. The whole process of the filament activation and eruption is nicely and sequentially portrayed in the zoomed cut-outs of Figures 6(d), (e), (f), and (h) (the filament is indicated by green arrows in these panels). It is noteworthy that the flare reached its peak intensity earlier in 304 \AA\ than in 94 \AA\ emission by $\sim$4 minutes (cf. Figure 1(b)). Following the eruption of the filament, we observed formation of two flare ribbons in the core region of the sigmoid (Figure 6(i); indicated by the white arrows in the zoomed cut-out). The separation between these two inner flare ribbons increased slowly but continuously (cf. Figures 6(i) and (l)) and after $\sim$12:14 UT, the loops connecting these two flares ribbons became so bright that they outshone the ribbons. Eruption of plasma from the elbow part of the sigmoid was also observed in the 304 \AA\ channel which is shown in the dotted boxes in Figures 6(j), (k), and (m) (also by the arrow in Figure 6(m)).

The late decay phase of the flare (after $\sim$12:32 UT) is characterized by very interesting evolution of highly extended flare ribbons that encompassed the whole active region (marked by the black arrows in Figures 6(n)--(o)). In Figure 6(o), we outline these ribbons by white dotted lines. Interestingly, the development of extended flare ribbons were spatially and temporarily correlated with the formation of hot coronal loop arcade, observed in AIA 94 \AA\ channel (cf. Figures 5(j) and (k)). Furthermore, the flux enhancement associated with evolution of extended flare ribbons and overlying loop arcade can be readily seen from EUV time profiles between $\sim$12:30 UT and $\sim$13:25 UT. It is interesting to note loop arcade in AIA 304 \AA\ images as well which essentially imply filling of plasma in low-lying loops at chromospheric temperature. Further, we note, very intense and diffuse emission from the loop arcade was observed in AIA 94 \AA\ channel while AIA 304 \AA\ images present much structured loop morphology (indicated by blue arrows in Figure 6(p)).

\begin{figure}
\epsscale{1.15}
\plotone{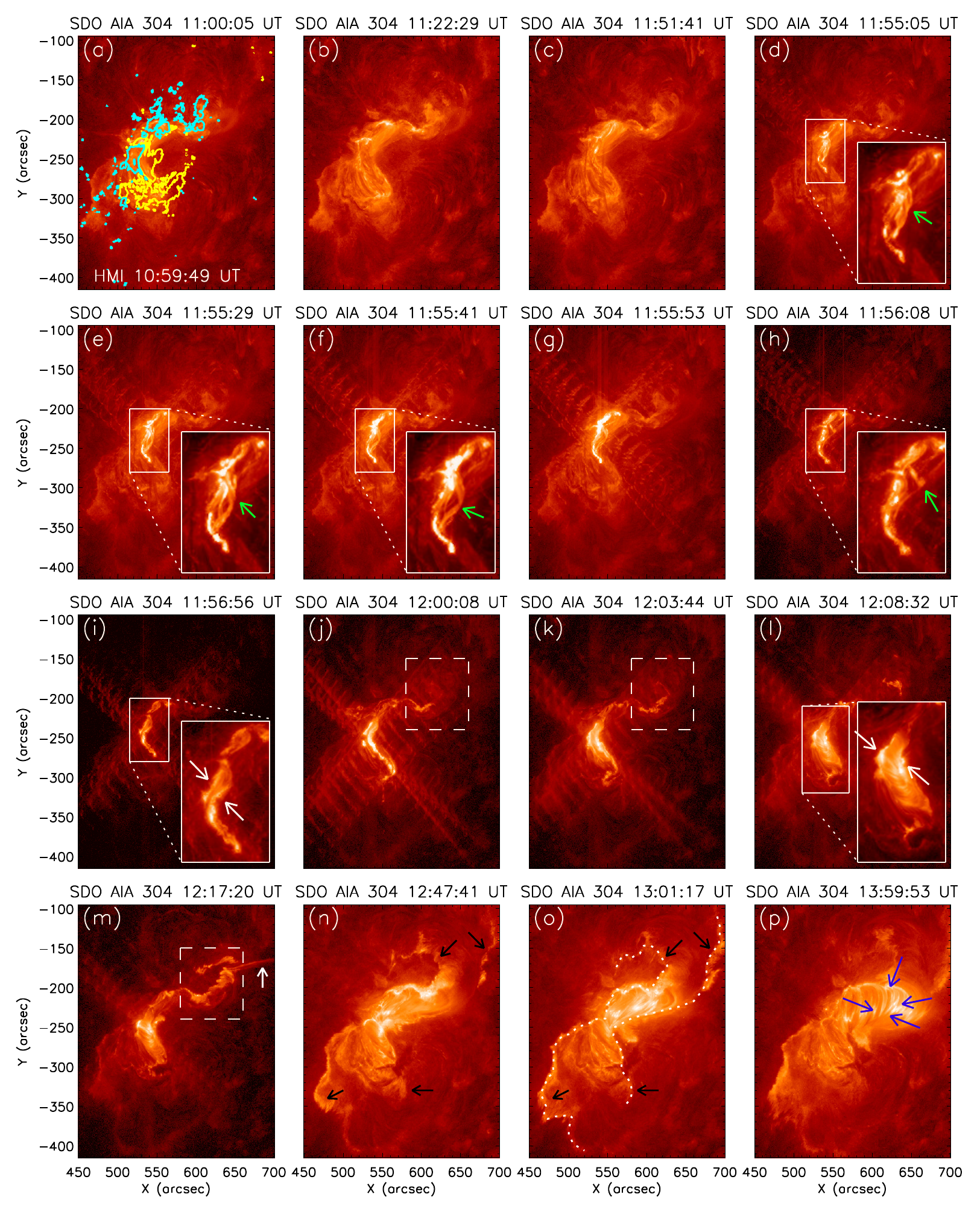}
\caption{Sequence of AIA 304 \AA\ images corresponding to the X9.3 flare. Insets in panels (d), (e), (f), and (h) sequentially show activation and eruption of the filament (indicated by green arrows). Insets in panel (i) and (l) show formation and separation of the two inner region post-flare ribbons (indicated by the two white arrows in the insets). Regions inside dashed boxes in panels (j), (k), and (m) show the semi-circular arc formed during the gradual phase of the flare. The small arrow in panel (m) indicate erupting plasma from the semi-circular arc region. Formation of extended post-flare loops in the decay phase of the X9.3 flare is shown by black arrows in panel (n) and (o), and by the dashed curve lines in panel (o). Post-flare arcades in the late decay phase are very clear in panel (p) (indicated by blue arrows). Contour levels of the HMI LOS magnetogram in panel (a) are ${\pm400\ G}$ where yellow and blue contours refer to positive and negative polarity, respectively.}
\end{figure}
\subsubsection{Plasma Eruption}

Spectacular eruption of plasma was observed in different phases during the X9.3 flare. Immediately after the eruption of the filament at $\sim$11:56 UT, we observed plasma eruption at two predominant directions: southern (shown in Figures 4(f)--(h) by yellow arrows) and north-western (indicated by blue arrows in Figures 4(f)--(i)). This first phase of the eruption ceased to exist after $\sim$12:03 UT in AIA 94 \AA\ difference images (cf. Figure 4(j)). However, a second stage of plasma eruption started to appear after $\sim$12:08 UT which continued until $\sim$12:55 UT while the flare had already entered in the late decay phase (indicated by red arrows in Figures 4(m)--(p)). As evident from the running difference images, this eruption mostly proceeded toward north-western direction. Here we recall reorganization occurring near the elbow region of the sigmoid following the filament eruption. Apart from these prominent phases of plasma eruption noted above, we cannot ignore continuous plasma eruption from the sigmoidal region during the flare that exhibited relatively faint signatures in the running difference images.

\subsection{DEM Analysis}
We have thoroughly examined spatial distribution and structures of multithermal plasma during different phases in the evolution of the sigmoid by adopting Differential Emission Measure (DEM) analysis. For this purpose, we utilize simultaneous observations in six coronal AIA channels (94 \AA , 131 \AA , 171 \AA , 193 \AA , 211 \AA , and 335 \AA). DEM, in the unit of cm$^{-5}$ K$^{-1}$, is given by
\begin{equation}
\label{1}
DEM(T) = n^2\frac{dh}{dT}
\end{equation}
where $n(h(T))$ is the electron density at height $h$ and temperature $T$ \citep[see Chapter 4 of][]{Mariska1992}. Line intensity, in unit of erg cm$^{-2}$ s$^{-1}$ sr$^{-1}$, which can be directly measured by the telescope detectors, is given by
\begin{equation}
\label{2}
I_{ji} = h\nu_{ji}\int G(T)\ DEM(T)\ dT + \delta I_{ji}
\end{equation}
where $G(T)$ is the temperature response function and j,i are atomic energy levels, i.e., the emission line is produced by the photons emitted due to the transition from $j^{th}$ energy level to $i^{th}$ energy level. $\delta I_{ji}$ is uncertainty in measurement. Total emission measure in a given temperature range $[T_1, T_2]$ can be expressed as
\begin{equation}
\label{3}
EM = \int_{T_1}^{T_2} DEM(T)\ dT
\end{equation}

The AIA temperature response function can be found by using SSW routine \textit{aia\_get\_response.pro}. For DEM computation, we have used the regularization inversion technique developed by \citet{Hannah2012}. The initial step to the process is to obtain perfect coalignment of AIA images in six coronal channels which can be done by SSW routine \textit{aia\_coalign\_test.pro}. This routine produces outstandingly accurate result (error being $<$1 pixel) by fitting limbs and makes the relevant changes in the header of the FITS file corresponding to the images \citep{Aschwanden2013}.

\begin{figure}
\epsscale{1.18}
\plotone{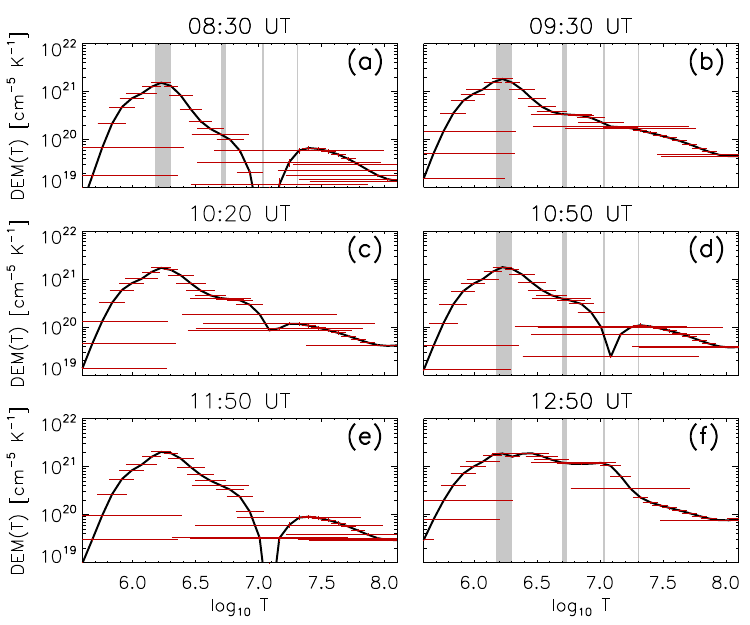}
\caption{DEM profiles obtained from the overall active region at six selected times: before the X2.2 flare (panel (a)), after the impulsive phase of the X2.2 flare (panel (b)), between the two X class flares (panel (c) and (d)), just before the initiation of the X9.3 flare (panel (e)), and after the impulsive phase of the X9.3 flare (panel (f)). The horizontal and vertical red colored line segments show uncertainty in temperature and DEM counts, respectively. Note that the vertical red lines showing uncertainty in DEM counts are very small compared to the horizontal lines because of the much larger range in DEM values (shown in the Y-axis) compared to the temperature values (shown in the X-axis). The gray-colored strips in panels (a), (b), (d), and (f) indicate four selected temperature ranges for which representative EM maps are given in Figure 8.}
\end{figure}
\begin{figure}
\epsscale{1.15}
\plotone{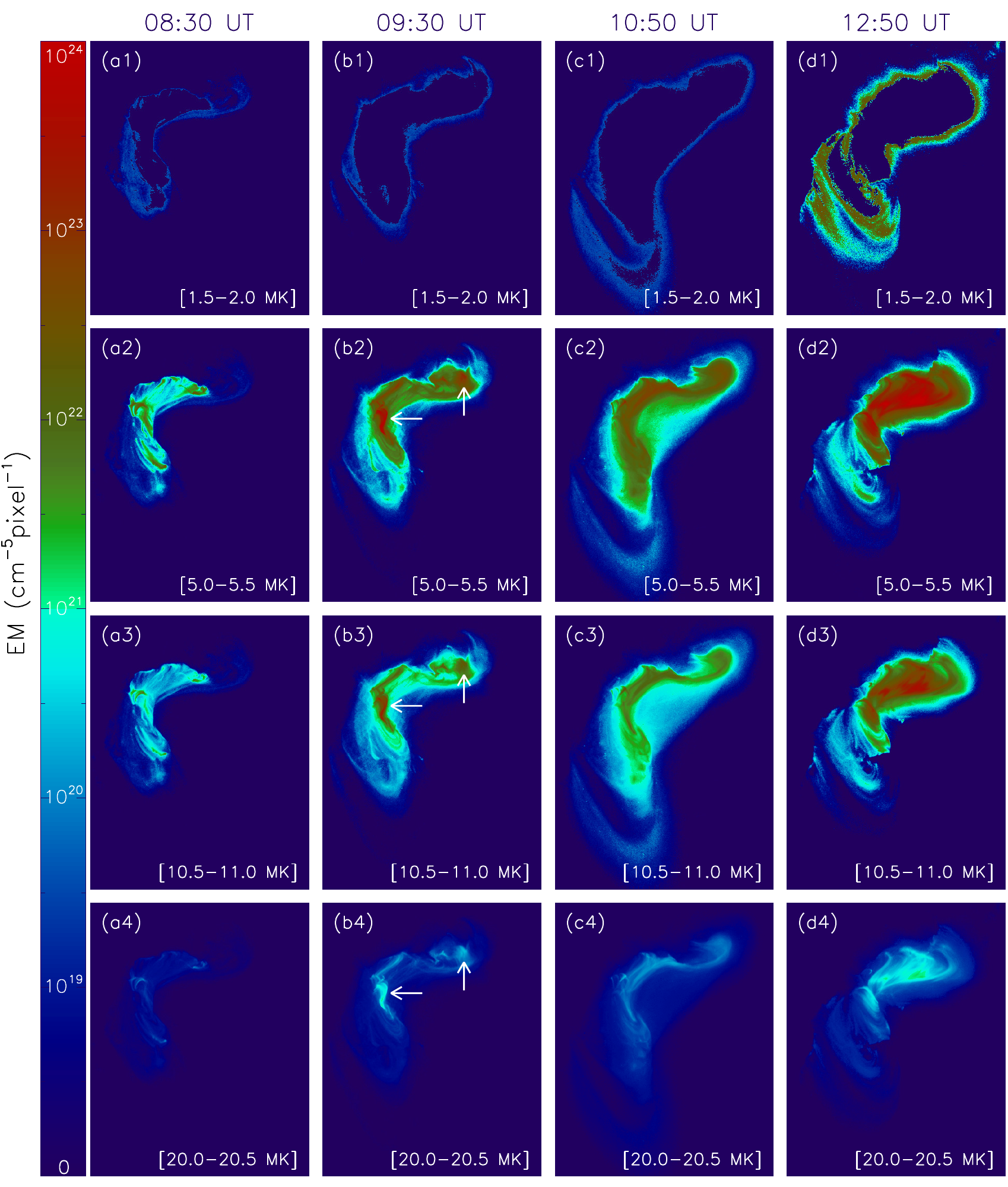}
\caption{EM maps of the active region in different temperature ranges 1.5--2.0 MK in row (a1)--(d1), 5.0--5.5 MK in row (a2)--(d2), 10.5--11.0 MK in row (a3)--(d3), 20.0--20.5 MK in row (a4)--(d4) before the X2.2 flare (column (a1)--(a4)), just after the X2.2 flare (column (b1)--(b4)), between the two flares (column (c1)--(c4)) and after the X9.3 flare (column (d1)--(d4)). The FOV of the EM maps is $(x_{1},y_{1})=(450'',-440'')$ ; $(x_{2},y_{2})=(700'',-150'')$. The four temperature ranges in these EM maps have been shown by shaded vertical strips in each of the panels (a), (b), (d), and (f) of Figure 7. In the panels of the first three rows, only those pixels with uncertainty both in DEM count and temperature less than 30\% , are plotted. In the panels of the bottom row, pixels with uncertainties in DEM count less than 30\% and temperature less than 50\% are plotted.}
\end{figure}
Figure 7 shows DEM count from the active region as a function of temperature at different times. Before the X2.2 flare (Figure 7(a)), we find two peaks in the DEM curve: one around log(T)$\approx$ 6.3 and another one around log(T)$\approx$ 7.4. Between log(T)$\approx$ 6.9 and 7.2, DEM counts are very low. After the impulsive phase of the X2.2 flare (Figure 7(b)), DEM counts in this temperature range enhanced significantly and thereafter decreased continuously till the start of the X9.3 flare (Figures 7(b)--(e)). Following the impulsive phase of the X9.3 flare, DEM counts in this temperature range enhanced again (Figure 7(f)). In Figure (7), we also show uncertainties in temperature and DEM counts by horizontal and vertical red colored lines. We note, uncertainty in temperature to be higher at low and very high temperature ranges (see temperature below log(T)=5.8 and above log(T)=7.8 in Figure 7(a)). However, the uncertainties at high temperatures reduced significantly just after the impulsive phases of the two X-class flares because of large enhancement of the overall emission of hot flaring loops from the active region.

For further understanding of plasma emission from the flaring region, we have plotted EM maps at four selected times (Figure 8) in different temperature ranges (indicated by the vertical strips in Figures 7(a), (b), (d), and (f)). The first column of Figure 8 (panels (a1)--(a4)) represents emission maps corresponding to the quiet phase of the active region (i.e., before the X2.2 flare). At lower temperature range (1.5--2.0 MK; Figures 8(a1)--(d1)), emission is seen to be predominantly produced from the outer regions of the sigmoid. This is acceptable as emission from quiet corona lies in this temperature range. We find that moderate amount of hot plasma existed in the sigmoid region before the flaring activities (Figures 8(a2) and (a3)). During the X2.2 flare, we find enhancement of emission measure at the core and northern elbow region (Figures 8(b2)--(b4)). Before the onset of X9.3 flare, the sigmoid underwent significant spatial expansion (Figures 8(c2)--(c4)). During the X9.3 flare, we note large enhancement in the emission measure of hot plasma at the northern part of the sigmoid while the southern part was highly structured.

\subsection{Magnetic Structure of AR 12673}

In Figure 9(a), an HMI LOS image of the active region is plotted. In the northern part of the active region, we find negative polarity magnetic regions predominantly existed while positive polarity structures were very less and dispersed. In the southern part of the active region, positive polarity regions strongly dominated over negative polarity regions. In the middle portion of the active region, we find strong regions of negative and positive polarities in a conjugated state. This was the most eventful region, being associated with activation and eruption of the filament along with early flare ribbon brightenings during the X-class flares (cf. Figure 9(b); also seen in Figures 3, 5, and 6). The blue box in Figure 9(a) encloses the photospheric region which was associated with inner ribbon brightening. In Figure 9(b), we present an overplot of AIA 304 \AA\ image with HMI LOS magnetogram to show the extension of flare ribbons with respect to the photospheric magnetic configuration during X9.3 flare. To estimate the magnetic field gradient across the PIL, we have considered a slit across the active region (the red line in Figure 9(a)) along which magnetic field strength and magnetic gradient were calculated which are plotted in Figure 9(c). We find a very sharp gradient in magnetic field with magnetic strength changing from about $\sim$ --1000 G to $\sim$ +1000 G over a distance of $\sim$1 arcsec (the peak gradient being $\sim2.4\times10^{3}$ G Mm$^{-1}$ on the PIL). In Figure 9(d), a continuum image of the active region is plotted, in which multiple, segmented umbrae can be found in the central region of the active region surrounded by a single penumbra. Comparison of the magnetogram and continuum images of the active region confirms that, prior to the X-class flares, AR 12673 had evolved into a complex `$\delta$-spot' type configuration.
\begin{figure}
\epsscale{1.1}
\plotone{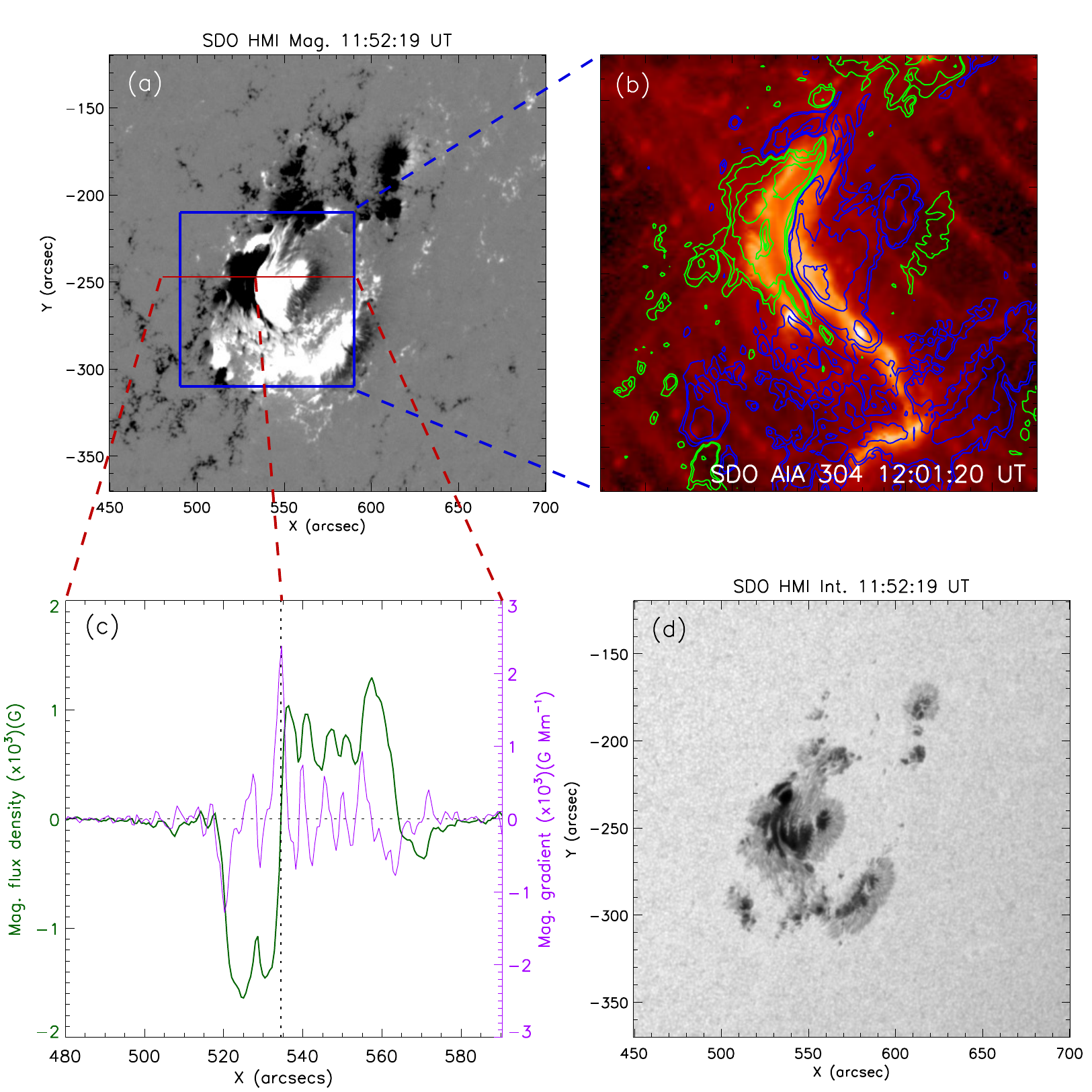}
\caption{Panel (a): HMI LOS magnetogram of AR 12673. The X-class flares occurred in the central region of the AR which is indicated by the blue box. Panel (b): AIA 304 \AA\ image of region showing the extent of inner flare ribbons during the impulsive phase of the X9.3 flare with contours from the LOS magnetogram overplotted. Contour levels are $\pm(300,500,1000,1500)$ G with blue and green contours referring to positive and negative polarity, respectively. Panel (c): Magnetic flux density (green) and magnetic gradient (purple) computed along the red slit in panel (a). Panel (d): White light image of the active region showing multiple, fragmented umbrae within conjugated penumbrae.}
\end{figure}

\section{Magnetic field modeling}

To understand the magnetic field topology of the active region, we extrapolate
magnetic field lines using the non-force-free-field (NFFF) technique \citep{Hu2008a, Hu2008b, Bhattacharyya2007}. For completeness, we 
provide a brief discussion of the technique in the Appendix.

\subsection{Pre-flare Configurations}
\begin{figure}
\epsscale{1.1}
\plotone{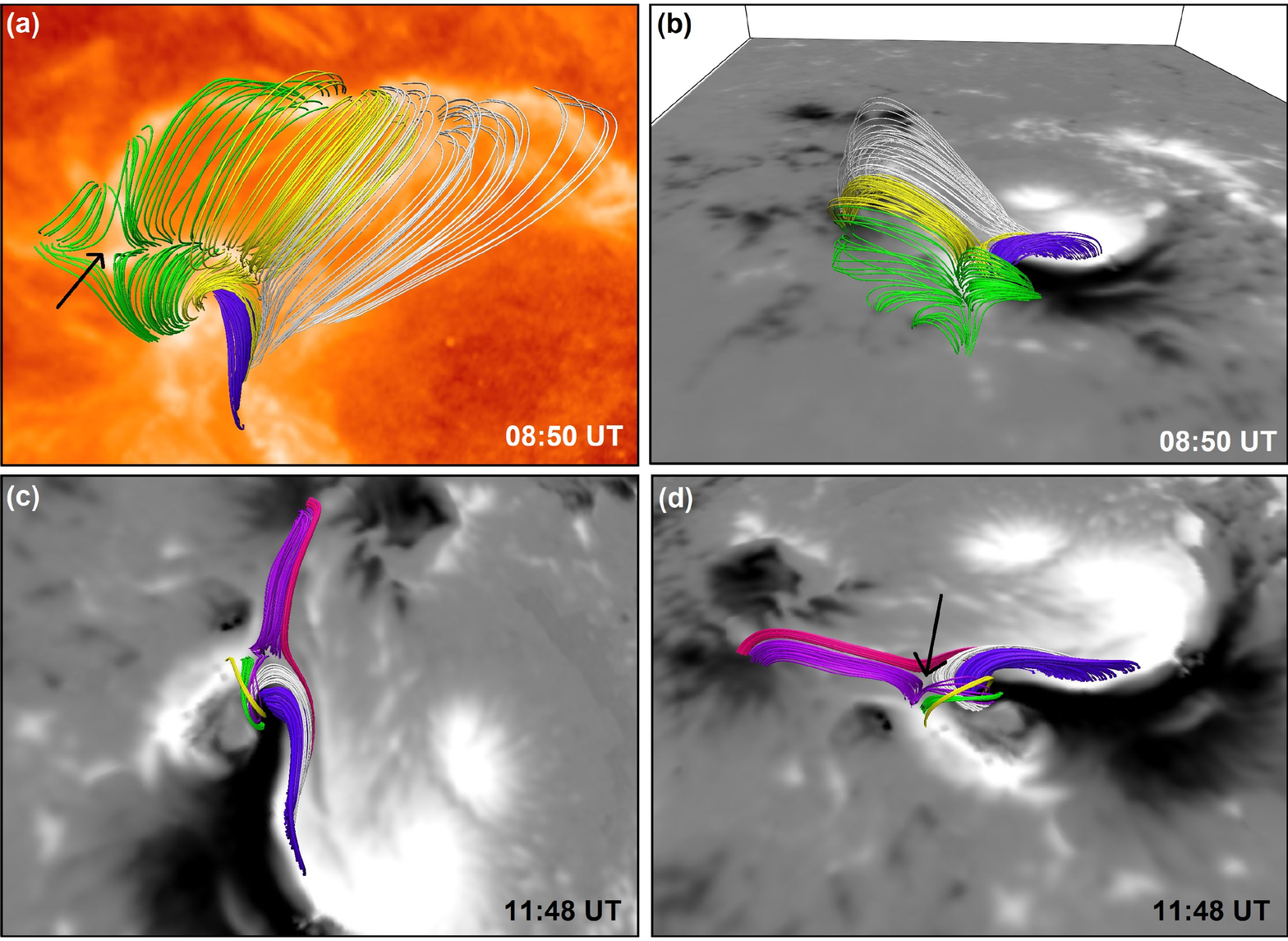}
\caption{NFFF extrapolation showing model coronal magnetic field structures of the active region. In panels (a) and (b), we show top and side views, respectively, of the coronal magnetic fields prior to the X2.2 flare (08:50 UT) associated with the core region of the sigmoid (i.e., high magnetic gradient region) and remote pre-flare ribbon like brightenings. In panels (c) and (d) we focus on the coronal magnetic configuration at the core region prior to the X9.3 flare (11:48 UT) with its top and side views, respectively. For clarity in representation, we have used different colors to show the model field lines. Background of panel (a) represents co-temporal AIA 304 \AA\ image whereas in panels (b)--(d), co-temporal HMI LOS magnetograms are plotted as background. We find a twisted set of field lines lying over the PIL, in the region of high magnetic field gradient (indicated by blue lines in panels (a)--(d)) that resembles with a magnetic flux rope. Black arrows in panels (a) and (d) indicate X-type null.}
\end{figure}

The modeled magnetic field configuration prior to the X2.2 flare (Figures 10(a) and (b)) reveals the pre-existence of a flux rope at the core of the sigmoid that extends over the PIL with a north-south orientation (shown in blue lines in Figures 10(a) and (b)) with northern and southern footpoints fixed at negative and positive polarity regions, respectively. Comparison of magnetic field extrapolations with pre-flare brightenings observed in AIA 304 \AA\ image (Figure 10(a)) reveals a very consistent picture in which ribbon like extended pre-flare brightenings seen in the north of the core region are formed at the footpoints of model active region loops that terminate at the core region (shown in green, yellow and white lines in Figures 10(a) and (b)). Further, extrapolated field lines near the northern footpoints of the flux rope reveal a magnetic null (indicated by the black arrow in Figure 10(a)) which is  regarded as a potential site for magnetic reconnection.

The overall pre-flare magnetic configuration, derived from the extrapolation, for the X9.3 flare at the core region of the sigmoid (Figures 10(c) and (d)) remains similar to that of the earlier X-class flare. Notably, the extrapolation shows the presence of a magnetic null (shown by the black arrow in Figure 10(d)) along with an X-type configuration (yellow and green lines in Figures 10(c) and (d)) at this time. Further, we find that field lines involved in the X-type configuration seem to provide tethering to the flux rope near its northern footpoint.

\subsection{Magnetic Configurations of the Inner and Extended Flare Ribbons During the X9.3 Flare}

As noted in section 3.3.1, the X9.3 flare is characterized by a pair of inner ribbons during the impulsive phase while extended ribbons appeared in the late phases (Figures 11(a) and (b)). In Figure 11, we show model coronal field configuration at the location and timing of the inner (12:08 UT) and extended (12:50 UT) flare ribbon structures. Extrapolations suggest the formation of a low-lying system of a highly sheared set of post-flare loops in the inner ribbon area (Figure 11(d)) which was occupied with the model flux rope during the pre-flare phase of the X9.3 flare (Figures 10(c) and (d)). Comparison between the extended ribbon structures during the late phase (12:50 UT) and extrapolated coronal field lines establish connectivities between opposite polarity ribbons by large overlying loop systems (Figure 11(e)). Connectivity in the northern part of the extended post-flare ribbons (shown in the box (I) in Figure 11(b)) is shown by the green colored model field lines in Figure 11(e), whereas, the southern part of the extended ribbons (shown in the box (II) in Figure 11(b)) are connected by the yellow colored model field lines (Figure 11(e)). Pink model lines in Figure 11(e) establish connectivities within the core region of the sigmoid in the late phase. Interestingly the model post-flare coronal loops at the central region of the sigmoid (blue lines) nicely resemble with the dense loop arcade observed in the AIA 304 \AA\ images (cf. Figures 11(c) and (e)).
\begin{figure}
\epsscale{1.1}
\plotone{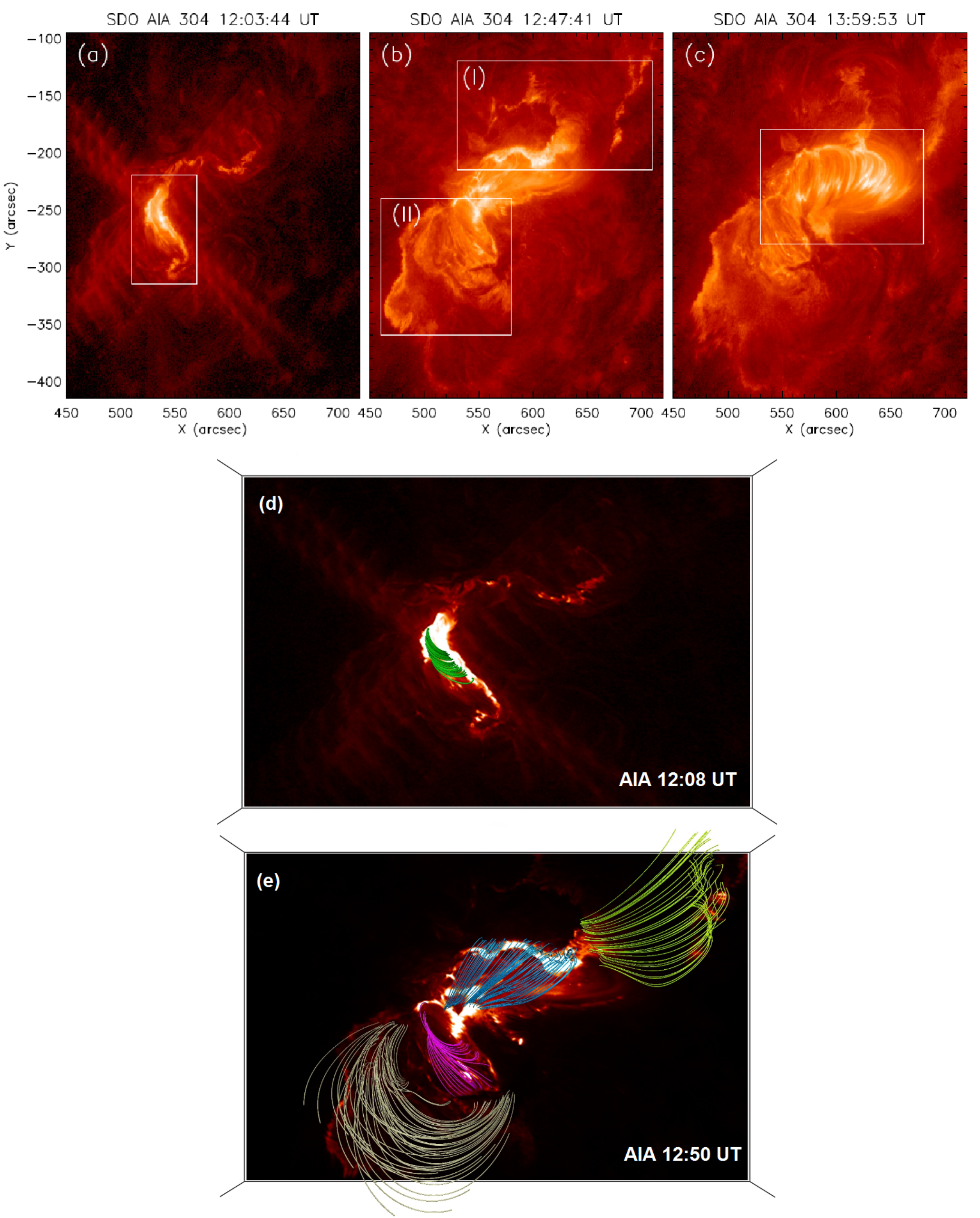}
\caption{Panels (a)--(c): AIA 304 \AA\ images showing the inner flare ribbons, extended flare ribbons, and post-flare loop arcades, respectively. Panel (d): Model coronal magnetic field configuration showing connectivity between the inner flare ribbons (also indicated by the box in panel (a)). Panel (e): Model coronal field configuration showing connectivities among different parts of extended ribbon structures by green and yellow lines (cf., regions marked by box (I) and (II)) in panel (b) and in the region of dense post-flare arcade by blue lines (also indicated by the box in panel (c)). Pink model lines in panel (e) show connectivity in the core of the active region in the late decay phase of the X9.3 flare.}
\end{figure}

\subsection{Free magnetic energy in the active region}
The magnetic free energy ($E_F$) associated with an active region can be estimated by the difference of non-potential magnetic energy (in this case, NFFF magnetic energy (${E_{NFFF}}$)) and potential magnetic energy (${E_P}$) i.e., $E_F=E_{NFFF}-E_P$. In Figure 12, we show variation of the normalized magnetic free energy (${E_F/E_P}$) for the extrapolated field from 08:30 UT to 14:00 UT which includes the two X-class flares. The shaded regions indicate the main energy release phases of the two X-class flares as identified in the GOES SXR band of 1--8 \AA\ (shown by the blue curve in Figure 12). Clearly the free energy of the active region decreased after both the X-class flares. As expected, the decrease in free energy observed was larger after the X9.3 flare than the earlier X2.2 flare.

We note that the free energy available in the active region prior to the eruptions was $\sim$82\% and $\sim$80\% of the potential energy for X2.2 and X9.3 flares, respectively. These estimates seem to be higher than typical active regions producing X-class flares. Several assessments concerning X-class events reveal the free energy to be $\sim$30--50\% of the potential energy \citep{Jing2010, Sun2012, Choudhary2013, Muhamad2018}. Although the availability of free magnetic energy is unambiguously a pre-requisite for flare occurrence, it alone may not ensure the flare onset \citep[see e.g.,][]{Jing2010}. We also emphasize that NOAA AR 12673 was an extraordinary active region that produced 4 X-class and 20 M-class flares within 6 days, while exhibiting a $\delta$-type magnetic configuration. Further, on 2017 September 6, two X-class flares (studied here) occurred very close in time (within $\sim$3 hours) with the latter one being the largest flare of the solar cycle, thus signifying the large excess of free magnetic energy of the AR 12673.

\begin{figure}
\epsscale{1.1}
\plotone{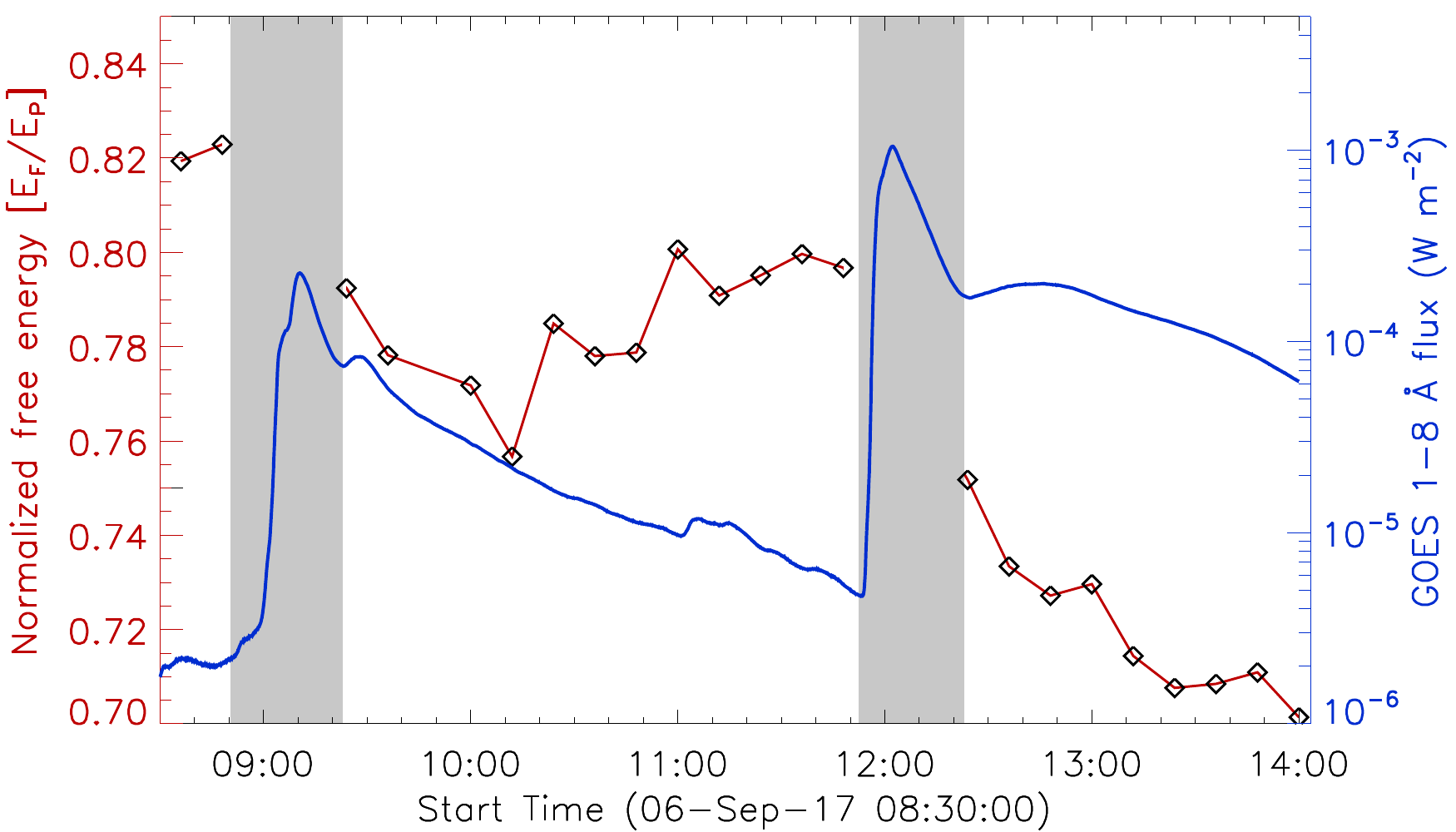}
\caption{Time variation of normalized free magnetic energy in the active region NOAA 12673 (shown by the red curve) on 2017 September 6 from 08:30 UT to 14:00 UT. Normalized free energy is computed by taking the difference between NFFF magnetic energy and potential magnetic energy, normalized by the corresponding potential magnetic energy (i.e., ${\frac{E_F}{E_P}}$). For a comparison, GOES 1--8 \AA\ flux profile in the same duration is shown by the blue curve which indicates the occurrence of two X-class flares. The shaded regions indicate the main energy release phase of the two X-class flares during which the intense flare emission caused artifacts in the measurement of the photospheric magnetic field making the calculations of magnetic energy unreliable.}
\end{figure}

\section{Discussion}
In this paper, we present multi-wavelength investigation of two X-class flares from the active region NOAA 12673 that occurred withing an interval of about 3 hours (cf. Figure 1). Notably, the second flare of X9.3 class turned out to be the largest flare in solar cycle 24. Both flares were eruptive in nature (Figures 2 and 4).

The active region NOAA 12673 was highly flare productive\footnote{\url{https://www.swpc.noaa.gov/products/solar-region-summary}}. It appeared in the eastern limb of the Sun on August 28 as a simple $\alpha$ type active region and gradually evolved into complex $\beta\gamma$ type on September 4. It became an even more complex $\beta\gamma\delta$ type on September 5 and remained so until its disappearance over the western limb of the Sun on September 10. In total, it produced 27 M-class and 4 X-class flares between September 4--10. Following the occurrence of the two X-class flares on September 6, reported in this paper, it went on to produce X1.3 class flare on September 7 and X8.2 class flare on September 10 besides several M-class major eruptive events. During the occurrence of X-class flares, the complex active region had shown ``$\delta$-sunspots" which are identified with a complex distribution of sunspot groups in which the umbrae of positive and negative polarities share a common penumbra \citep{Kunzel1960}. Such complex active regions are known to produce powerful flares \citep[see e.g.,][]{Zirin1987, Takizawa2015, Sammis2000}. It is noteworthy that AR 12673 was a rather compact region (Figure 9) which displayed more spatial extension in the north-south direction than the usual east-west direction. The $\delta$-sunspots were concentrated at the central part of the active region where magnetic fields were very strong and magnetic field gradient across the PIL was extremely high ($\sim2.4\times10^3$ G Mm$^{-1}$; see Figure 9(c)). Earlier studies have shown a close relation between major flare activities and strong magnetic field; specially those with high gradient and highly sheared across the PIL \citep{Hagyard1984, Zirin1993, Schrijver2007, Barnes2016}. The reported eruptive activities in AR 12673, thus, represent the  capability of the AR in the rapid generation and storage of huge amount of excess magnetic energy in the corona. In this context, the evolution of normalized free magnetic energy during the X-class flares is noteworthy (Figure 12). We find that the free magnetic energy stored in the AR before the flaring activity was $\sim$82$\%$ of the potential magnetic energy. After the two X-class flares, it reduced to $\sim$70$\%$. Our analysis, therefore, implies that a large amount of free magnetic energy was already stored in the active region before the flaring activities and the large X-class flares essentially released only a  small fraction of it.

The observations of the AR in the AIA 94 \AA\ channel (6 MK) clearly show the pre-existence of a coronal sigmoid which is considered to be an indicator of twisted or helical magnetic structures \citep{Gibson2002}. Coronal sigmoids, originally discovered in SXR emission \citep{Manoharan1996, Rust1996}, were eventually identified as potential sites for CMEs \citep{Canfield1999}. Later on, sigmoid type active regions were confirmed in numerous EUV observations \citep{Sterling2000, LiuC2007, LiuR2010, Joshi2017, Green2018, Joshi2018}. Following the eruption of the flux rope, sigmoid structures are observed to undergo major reorganization and the region is enveloped by a set of bright post-flare loop arcades or cusped loops. Thus, the ``sigmoid-to-arcade" development is suggestive of large-scale magnetic reconnection driven by the eruption of the sheared core fields \citep{Sterling2000}. Our observations reveal sigmoid structure prior to the first X-class flare which remained intact following the event (Figure 3). However, the sigmoidal region completely transformed into the post-flare arcade in a sequential manner during the second X-class flare (Figures 5 and 6). The present observations thus indicate the occurrence of two eruptive X-class flares within a single ``sigmoid-to-arcade" event, which does not seem to be a commonly observed phenomenon. In a comprehensive study of a major geomagnetic effect caused by two ultra-fast CMEs  in 2012 March, \citet{Patsourakos2016} found that the two CMEs were linked with two X-class flares occurring from the same active region (NOAA 11429) within an interval of $\sim$1 hour. However, unlike our studied events, the two X-class flares were initiated from different locations of the active region and were associated with two separate sets of post-flare arcades.

Soon after the eruption of the flux rope from the central portion of the active region during the impulsive phase of the X9.3 flare, formation of two flare ribbons on the either side of the PIL was observed in AIA 304 \AA\ images (the inner flare ribbons; see Figure 6(i)). We note that the flare ribbons developed during the previous X2.2 event also appeared at the same part of the active region, i.e., on either side of the high magnetic gradient region along the PIL (Figure 3(j)). It is noteworthy that these inner flare ribbons were subjected to very little separation which is rather uncommon in strong eruptive X-class flares. We attribute this small separation between flare ribbons to the fact that footpoints of the flare loops were rooted in $\delta$-sunspot regions of strong magnetic field $\mathbf{B}$. Thus, although the loop growth and inner ribbon expansion speed $\mathbf{v}$ were small, the associated local reconnection rate (coronal electric field, i.e., the rate at which field is brought into the reconnection region) which can be derived from the observations of the flare ribbons as $\mathbf{E_c}=\mathbf{v}\times\mathbf{B}$ \citep{Forbes1984} would have been large enough to produce X-class events. This is supported by the finding of the recent statistical analysis of reconnection rates in solar flares over 4 orders of flare magnitude (from B to X17) by \citet{Hinterreiter2018}, who showed that the correlation of the local reconnection rate with flare class is a stronger function of the underlying field strength $\mathbf{B}$ than the flare ribbon separation velocity $\mathbf{v}$ (see Figure 10 therein).

We performed DEM analysis to probe various temperature structures of the sigmoid (Figure 8). Our analysis reveals that the temperature distribution of the sigmoid was much structured. Interestingly, the core of the sigmoid is observed to produce high temperature emission, even prior to the flare activities, while the outer regions of the sigmoid were seen at lower temperatures. The pre-flare heating of the sigmoid is likely due to the localized events of small-scale, slow reconnections that may occur in response to the flux cancellation at the PIL \citep{Moore1992}. The core of the sigmoid expanded during the X-class flares which is attributed to strong heating by large-scale magnetic reconnections. We further note that the sigmoid region was already much heated prior to the X9.3 flare. This pre-heating of plasma could significantly contribute toward enhancing the efficiency of the particle acceleration process associated with subsequent flare emission. EM maps also reveal the substantial volume of hot ($>$10 MK) plasma during the post-flare phase of the X9.3 event.

NFFF modeling of coronal magnetic field lines clearly show a twisted set of magnetic field lines at the core of the sigmoid which can be clearly identified as a flux rope \citep{Cheng2014}. We note intense short loops that seem to tether the middle region of the sigmoid (Figure 3(b)). Notably, this region intensified highly during the impulsive phase of the flares (Figures 3(c)--(f) and Figures 5(e)--(g)). We identify this as the ``core" of the sigmoid which lay over the region of very high magnetic gradient and high non-potentiality. The tether of the flux rope is apparently enveloped by the short, low-lying magnetic loops (Figure 10) which suggest a close match between observed and modeled coronal loops. Our magnetic extrapolation analysis also successfully reveals magnetic connectivity between highly sheared core fields, close to the flux rope, and pre-flare ribbon like brightenings observed in the 304 \AA\ channel (Figure 10(a)).

We emphasize that both the X-class events not only occurred at the same region but also share a common initiation process. The `trigger' happened at the central part of the sigmoid (Figures 3(c)--(e) and Figures 5(e)--(g)) which spatially lay close to the northern edge of the high magnetic gradient area (Figure 9) and was subjected to intense heating subsequently (Figure 8). The detection of pre-existing magnetic null, close to the northern edge of the flux rope structure, in the extrapolated field lines (Figure 10) have important implications in understanding the triggering mechanism for the subsequent eruptive events. The synthesis of observed and modeled coronal loops/field lines provide concrete evidence that the location of the initial energy release is spatially correlated with the site of the magnetic null. In several contemporary studies on solar eruptions, it is widely accepted that 3-dimensional magnetic null provides a favorable configuration for magnetic reconnections. We thus believe that the triggering mechanism for the eruptive events under this study are consistent with the breakout model \citep{Antiochos1999} of solar eruptions. The observations of pre-flare ribbon structures which were magnetically connected to the core region (Figure 10(a)) further strengthens the case for the breakout scenario as the overlying field lines would channelize the flow of particles, accelerated in the breakout current sheet, to remote footpoints. However, unlike the case of the classical breakout model, here pre-flare reconnection seems to have occurred at relatively lower coronal heights. Our observations indicate that the `triggering reconnections' at the magnetic null were capable of destabilizing the flux rope only locally during the first X-class flare as the eruption was relatively minor and the overall sigmoid structure was preserved. This could happen due to the strong field lines lying over the flux rope as well as firm tying of its southern footpoint. The coronal conditions seem to become conductive for the complete eruption by the time of the second X-class flare which resulted in the eruptive phenomena at much larger scales. Notably, the X9.3 flare exhibited two well-separated sets of large flare ribbons, the inner and outer ones, during the impulsive and declining phases, respectively, which imply that the reconnection in the overlying loops during the respective stages progressed at different heights in the corona. The brightening within the inner flare ribbon area marked the peak phase of this large X9.3 flare. On the other hand, the growth of the outer flare ribbons was accompanied by a complete restructuring of the sigmoidal region with the successive formation of a post-flare arcade. Thus, the evolution of the inner and the outer flare ribbons signify the role of large-scale magnetic reconnection in this long duration event.

As the flares progressed, the AIA 94 \AA\ images revealed outward moving arc-like high-temperature plasma structures which are termed as hot channels in contemporary literature \citep{Patsourakos2013,Cheng2013,Cheng2016, Nindos2015, Joshi2017}. The AIA 94 \AA\ running difference images unambiguously detect a hot channel eruption during both X-class flares (Figure 4) which is particularly prominent for the second event. The EUV hot channels have been accepted as evidence of magnetic flux ropes and, therefore, perhaps the earliest signatures of a CME in low corona \citep[see e.g.,][]{Patsourakos2013,Cheng2013, Joshi2017} which is also supported by our observations.

In summary, our detailed study of two X-class flares from AR 12673, within a single ``sigmoid-to-arcade" event, has provided important insights into the processes that are associated with the triggering of flux rope eruption and subsequent energy release processes. We find the inner region of the sigmoid to be at much higher temperatures compared to its outer regions and it also showed structured plasma emission even during the pre-flare phase. During the first X-class flare the core region of the sigmoid attained very high plasma temperatures which persisted afterward while during the second flare the intense brightenings at the core further extended outward, suggesting acceleration of the electrons and subsequent energy deposition by them occurred at much higher scales during the second event. The eruption of the hot EUV plasma channel during the flares and their subsequent expansion and outward propagation provide evidence of flux rope eruptions as the earliest signatures of CME from the source region. The pre-existence of the flux ropes at the high magnetic field gradient region at the core of the sigmoid, identified in the NFFF modeling of the coronal magnetic field, is well supported by the subsequently observed eruptive expansion of hot EUV plasma channels which also originated from the core region. Our observations support the breakout model of solar eruptions. The second X-class flare diverged from the standard flare model in the evolution of two sets of flare ribbons that are spatially well separated, inner and outer ones, providing firm evidence of magnetic reconnections at two coronal heights involving low-lying and higher coronal loop systems, respectively. We conclude that the very high magnetic gradient across the PIL and excessive storage of non-potential energy in the active region were the main cause behind the repeated strong flaring activity in the $\delta$-sunspot configuration of AR 12673.

\acknowledgments
We thank the SDO team for their open data
policy. SDO is NASA's missions under the
Living With a Star (LWS) program. We are also thankful to Q. Hu for sharing the NFFF extrapolation code with us. This work is supported by the Indo-Austrian joint research project no. INT/AUSTRIA/BMWF/
P-05/2017 and OeAD project no. IN 03/2017. A.M.V also thanks the Austrian Science Fund (FWF): P27292-N20. We acknowledge the constructive comments and suggestions of the anonymous referee that improved the scientific content and the presentation of the paper.

\appendix
\section{The Non-Force-Free extrapolation}
The advocacy of the NFFF extrapolation finds its root in the following dimensional analysis:
\begin{equation}
\label{forcebalance}
\frac{|\mathbf{j}\times \mathbf{B}|}{|\rho \frac{d\mathbf{v}}{dt}|}\sim\frac{B^2}{L}\frac{t}{\rho v}\sim\frac{B^2}{\rho v^2}\sim\frac{B^2}{\rho {v_{th}}^2}\frac{v_{th}^2}{v^2}\sim\frac{1}{\beta}\frac{v_{th}^2}{v^2}
\end{equation}
where ${\mathbf{v_{th}}}$ and $\mathbf{j}=\nabla\times
\mathbf{B}$ are respectively the thermal velocity and the volume current density. 
The speed of the photospheric flow is generally accepted to be $\sim$1 km s$^{-1}$ \citep{Vekstein2016,Khlystova2017} and a straight forward calculation finds the thermal speed of the photospheric plasma to be also $\sim$1 km s$^{-1}$. 
Therefore, with ${v_{th}}\sim v$ on the photosphere,

\begin{equation}
\frac{|\mathbf{j}\times \mathbf{B}|}{|\rho \frac{d\mathbf{v}}{dt}|}\sim\frac{1}{\beta}~~.
\label{lorentz}
\end{equation}
In the above equation, $\beta$ is the ratio of thermal to magnetic pressure
which can be of the order of unity on the photosphere under an equipartition 
of kinetic and magnetic energies. Equation (\ref{lorentz}) then  yields
\begin{equation}
|\mathbf{j}\times\mathbf{B}|\approx|\rho\frac{d\mathbf{v}}{dt}|~~,
\end{equation}
indicating the importance of the Lorentz force in conditions where $\beta\approx 1$.

The employed NFFF extrapolation uses an
inhomogeneous double-curl Beltrami
equation for the magnetic field $\mathbf{B}$,
\begin{equation}
\nabla\times\nabla\times\mathbf{B}+a\nabla\times\mathbf{B}+b\mathbf{B}=\nabla \psi
\label{e:bnff}
\end{equation}
\noindent where $a$ and $b$ are constants. The scalar function $\psi$ must be a solution of Laplace equation to make ${\bf{B}}$ solenoidal. A modified vector ${\bf{B}}^\prime={\bf{B}}-\nabla\psi$ \citep{Hu2008b} satisfies the corresponding homogeneous equation which, is known to represent a two-fluid MHD steady state \citep{Mahajan1998} having a solution
\begin{equation}
\mathbf{B}^\prime = \sum_{i=1,2} \mathbf{B_i}
\label{e:b123}
\end{equation}
where each $\mathbf{B}$ is a linear-force-free-field satisfying
\begin{equation}
\nabla\times\mathbf{B_i}=\alpha_i\mathbf{B_i}
\label{e:b124}
\end{equation}
in usual notations. The two sets of constants are related by
 $a=-(\alpha_1+\alpha_2)$ and $b=\alpha_1 \alpha_2$. The magnetic field is 
\begin{equation}
\mathbf{B} = \sum_{i=1,2} \mathbf{B_i}+\mathbf{B_3}
\label{e:b125} 
\end{equation}
where $\mathbf{B_3}=\nabla\psi$ is a potential field. 
Combining Equations (\ref{e:b124}) and (\ref{e:b125}),
\begin{align}
\begin{pmatrix}
\mathbf{B_1}\\
\mathbf{B_2}\\
\mathbf{B_3}\\
\end{pmatrix}
=\mathcal{V}^{-1}
\begin{pmatrix}
\mathbf{B}\\
\nabla\times\mathbf{B}\\
\nabla\times\nabla\times\mathbf{B}\\
\end{pmatrix}
\label{e:vmat}
\end{align}
where the matrix $\mathcal{V}$ is a Vandermonde matrix having elements $\alpha^{i-1}_j$ for $i, j = 1, 2, 3$ \citep{Hu2008a}.

The procedure of NFFF extrapolation, described above, requires at least two layers of vector magnetogram since it involves the solution of a second order derivative, $\nabla\times\nabla\times\mathbf{B}=-\nabla^2\mathbf{B}~$, at $z=0$. This imposes a severe limitation in view of the availability of only one single-layer vector magnetogram. Thus, the NFFF solution is not unique and other solutions could exist on the same boundary condition. Therefore, the results of magnetic field modeling based on NFFF extrapolation should be viewed under this limitation. In view of this situation, we follow the technique proposed and demonstrated in \citet{Hu2008b, Hu2010}. In brief, starting with a
selected pair of $\alpha$ and $B_3=0$, an optimal pair of $\alpha$'s
is obtained
by minimizing the average normalized
deviation of the magnetogram transverse field $(\mathbf{B_t}$) from its extrapolated value
$(\mathbf{b_t}$), namely
\begin{equation}
E_n =\sum_{i=1}^M |\mathbf{B}_{\mathbf{t},i}-\mathbf{b}_{\mathbf{t},i}|/\sum_{i=1}^M |\mathbf{B}_{\mathbf{t},i}|
\end{equation}
where $M$=$N^2$ is the total number of grid points on the transverse plane.
A fine tuning of ${E_n}$ is further ensured by using $\mathbf{B_3}=\nabla\psi$ as a 
corrector component field for the given pair of $\alpha$'s. Noting a superposition of 
potential fields being always potential, the above procedure is repeated until 
a minimal value of $E_n$ is asymptotically achieved. For the extrapolations reported here, 
Figure 13 shows the variation of $E_n$ with the number of iterations and documents its asymptotic approach to a minimal value. Auxiliary calculation (not shown) finds the average intensity of the extrapolated photospheric magnetic field, related and prior to the X9.3 flare, is $\approx${1.2} kG. The corresponding $\beta\approx0.97$ (for $ n\approx 10^{23}\ m^{-3}$ \citep[see,][]{Priest2004} and $T\approx 4000$ K, standard notations) substantiates the importance of the Lorentz force and makes the 
obtained extrapolated field, albeit non-unique, reliable.

Noteworthy is also the field line twist for the superposed ${\bf{B}}$. Twist can be quantified by the field-aligned current which, in this case, is given by
\begin{equation}
\label{e:twist}
\tau\equiv\frac{\mathbf{J}\cdot\mathbf{B}}{|\mathbf{B}|^2} = \frac{(\alpha_1\mathbf{B_1}+\alpha_2\mathbf{B_2})\cdot(\mathbf{B_1}+\mathbf{B_2}+\mathbf{B_3})}{(\mathbf{B_1}+\mathbf{B_2}+\mathbf{B_3})\cdot(\mathbf{B_1}+\mathbf{B_2}+\mathbf{B_3})}~~.
\end{equation}
In addition to any  modifications of $\alpha_1$ and $\alpha_2$, the twist $\tau$ can vary because of changes in the component fields which is a standout advantage of the NFFF extrapolations. To appreciate, we notice the $\alpha$'s to have an inherent upperbound ($\alpha_i\leq\frac{2\pi}{n}$; n is the dimension of the active region cut-out) which ensures a monotonous decay of the magnetic field with height \citep{Naka1972}. As the field line twist depends on $\mathbf{B_1}$, $\mathbf{B}_2$ and $\mathbf{B_3}$ in addition to $\alpha_1$ and $\alpha_2$, variations in them can accommodate any extra twist which would otherwise be impossible to accommodate because of the maximal limit in $\alpha$'s. An example is the active region analyzed here. The two $\alpha$'s have attained their maximal values at which they remained fixed before and after the flares--whereas post flare, the free energy decreases (see, Figure 10) because of readjustments of all component fields at the bottom boundary.
\begin{figure}
\epsscale{1.1}
\plotone{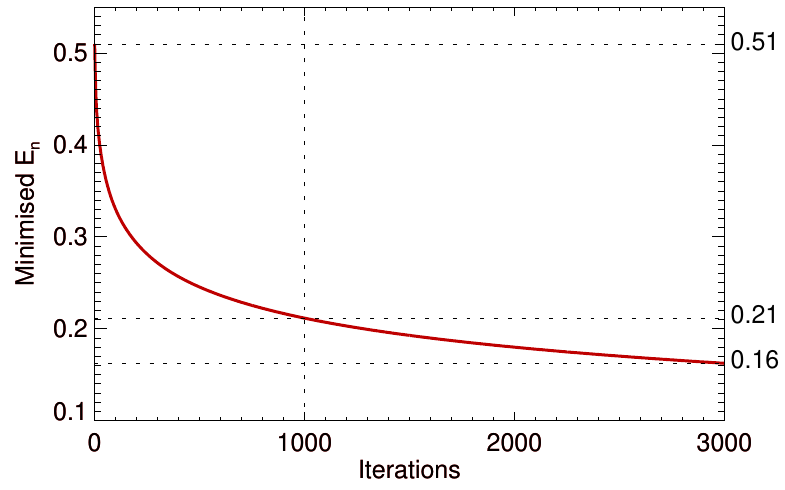}
\caption{Variation of minimized deviation ($E_n$) with number of iterations corresponding to NFFF extrapolation for the case of the active region NOAA 12673 on 2017 September 06. $E_n$ decreases monotonically to reach a value of 0.2 asymptotically for 1000 iterations. In order to save the computational cost, we have used the NFFF code for magnetic extrapolation up to 1000 iterations in this paper.}
\label{En}
\end{figure}

\end{document}